\documentstyle[preprint,aps]{revtex}
\documentstyle[12pt,epsf]{article}
\title{
Simultaneous Microscopic Description
of $^{16}O(\gamma ,n)$ and $^{16}O(\gamma ,p)$
Including $\Delta$ Degrees of Freedom}
\author{Thomas B. Bright and Stephen R. Cotanch \\
{\em North Carolina State University, Raleigh, N.C.  27695}}
\date{\today}
\begin{document}
\maketitle
\baselineskip=24.0pt

\begin{abstract}
The photonuclear reactions $^{16}O(\gamma ,n)^{15}O$ and
$^{16}O(\gamma ,p)^{15}N$
are simultaneously described within a coupled channels
framework based on a continuum shell model
formulation that includes all
non-localities arising from antisymmetrization.
This large-scale conventional
many-body calculation
provides a good description of the medium energy data.
The inclusion of $\Delta (1232)$ degrees of
freedom further, but slightly, improves the description of the available
data.
Complete quantitative agreement however is still lacking.
\end{abstract}

The apparent inability to describe
and understand fully photonuclear
reactions below 400 MeV
using conventional, non-relativistic dynamics
has attracted growing theoretical and experimental interest.
This theoretical shortcoming has motivated many calculations
that have separately investigated different
mechanisms thought
to be important, such as
nucleon-nucleon correlations~\cite{SCHO,GORI},
relativistic equations
of motion~\cite{MCDE,LOTZ,BOFF}, meson
exchange currents~\cite{GARI,CAVI,RYC1,RYC2},
and excitation of  $\Delta (1232)$
resonances~\cite{LOND,CHEU}.
Despite such efforts, the available $(\gamma ,p)$
data~\cite{LEIT,FIND,ADAM} below 400 MeV are
described
at best only semi-quantitatively,
while the computed
$(\gamma ,n)$ cross sections~\cite{BEIS}
are not even qualitatively described.
The purpose of this letter is to report
a large-scale, super-computer based
$\it {ab}$ $\it {initio}$
calculation that includes
both nucleon correlations and
$\Delta (1232)$ isobar excitations.
Our objective is to provide a realistic
microscopic framework within which to
consistently assess the importance of
the aforementioned effects.
We find that $\Delta (1232)$ isobar effects are small except at
very large momentum transfer, and that a
comprehensive conventional many-body structure calculation
is sufficient to obtain
the correct magnitude and
qualitative features for both
$(\gamma ,p)$ and $(\gamma ,n)$ reactions
over a wide range of energies and angles.

The calculation is a
generalization of the microscopic continuum
shell model
treatment originally developed
by Buck and Hill~\cite{BUCK}.
The many-body Hamiltonian for $^{16}O$ is
diagonalized rigorously in a model space
spanned by products
of continuum single-particle states and
the mass $A=15$ bound states.
We use a
realistic two-body, finite-range
nucleon-nucleon interaction with spin, isospin, and
tensor components,
and we rigorously
include all non-localities arising from antisymmetrization.
The central $N-N$ interaction
between nucleons i and j is
$$
{
V_c{ }\big [ a_o{ }  +a_\sigma (\vec  {\sigma _i}
\cdot \vec {\sigma _j})
{ }+{ }a_\tau (\vec {\tau _i} \cdot \vec {\tau _j})
{ }+{ }a_{\sigma \tau} (\vec {\sigma _i} \cdot \vec {\sigma _j})
(\vec {\tau _i} \cdot \vec {\tau _j})
\big ]
\big[ Y_o(\mu _c r)
{ }-{ }{\Lambda \over {\mu _c}} Y_o(\Lambda r)
\big ],
}
$$
while the tensor interaction has the form
$$
V_t{ }(\vec {\tau _i} \cdot \vec {\tau _j})
{ }S_{12}{ } \big [ Y_2 (\mu_t r)-
{({\Lambda \over {\mu _t}})}^3 Y_2(\Lambda r)\big ],
$$
where
\begin{eqnarray*}
Y_0(x){ }&=&{ }{e^{-x} \over x},  \\
Y_2(x){ }&=&{ }(1+{3 \over x} + {3\over {x^2}}) { }Y_o(x),
\end{eqnarray*}
and
$$
S_{12}{  }={  }3(\vec {\sigma _i} \cdot \hat r)
(\vec {\sigma _j} \cdot \hat r)
- (\vec {\sigma _i} \cdot \vec {\sigma _j}).
$$
The range parameters $\mu _c,\mu _t$ and the effective
strengths $V_c,V_t$ for the respective interactions were
previously
determined phenomenologically
by describing the giant dipole resonance for
$^{16}O(\gamma ,p)^{15}N$.
We
include $\Delta$ isobars by expanding the model space
to incorporate explicitly $\Delta $ components,
in addition to the
usual nucleon components, in the scattering wavefunction.
The form of the $N-\Delta $
and $\Delta -\Delta $ interaction potentials
is taken from the work of Niephaus, Gari, and Sommer~\cite{NIEP},
which
explicitly involves both $\pi$ and $\rho$ meson exchange.
The central $\Delta $ interaction Hamiltonian
has a radial form that is proportional to
$$
Y_o(\mu _c r)
{ }-{ }{\Lambda \over {\mu _c}} Y_o(\Lambda r)
\big [
1{ }+{{\Lambda ^2 -\mu _c ^2}\over
{2\Lambda ^2}
} \Lambda r\big ],
$$
while the tensor $\Delta $ interaction Hamiltonian is given by
$$
Y_2 (\mu_t r)-
{({\Lambda \over {\mu _t}})}^3 Y_2(\Lambda r)
-{\Lambda \over \mu _t}
{{{\Lambda ^2 -\mu _t ^2}\over
{2\Lambda ^2}}
} \Lambda r Y_1(\Lambda r)
,
$$
where
$$
Y_1(x){ }={ }(1+{1 \over x} ) { }Y_o(x).
$$
The quantity $\Lambda$,
introduced to regularize the Yukawa singularity
at r=0, is the same for all interactions.
The central interaction has only a
spin-isospin
term in the case of $N-\Delta $
transitions, whereas for the $\Delta -\Delta $ case,
the form is
similar to the $N-N$ interaction.
We assume that the allowed $\Delta $ excitations
are intermediate,
and hence do not include $\Delta $ components in the
$A=15$ bound states or the final state of the $A=16$ system.
Further, these many-body bound states are calculated within
the single-particle - single-hole limit of the shell model.
The resulting large set of coupled integro-differential
equations is solved numerically for the
continuum scattering wavefunction.
In order to perform the calculation,
the use of high-performance facilities is necessary, taking
advantage of their vectorization and run-time storage abilities.
For more specific details, the reader is referred to
references~\cite{HOI1,HOI2}.

The electromagnetic transition amplitude,
$
M=-\int d\vec r { }
\Psi _f { } \vec j(\vec r) \cdot \vec A(\vec r){ }
\Psi _i ,
$
entails the nuclear current $\vec j(\vec r)$
and the electromagnetic potential
$\vec A(\vec r)$,
where $\Psi_i$
is the initial nuclear state obtained from the
coupled channels calculation,
and
$\Psi_f$
is the final single-particle - single-hole
shell model wavefunction
describing the $^{16}O$ ground state.
The usual multipole expansion is
performed and, for the medium energies
considered here,
we found it necessary to include
electric multipoles through E17 and
magnetic multipoles through M15 to insure
convergence of the series.
The nucleon currents used in the present study are
the one-body convective and magnetization currents.
Because we do not allow $\Delta$ resonances
in the final nuclear state, the $\Delta$ current
involves only radiative transitions to the nucleon.
Work is in progress to include the
effects of two-body exchange currents.

Figure 1 summarizes our calculation
for the reaction
$^{16}O(\gamma ,p)^{15}N$
below $E_{\gamma }=400 MeV$.
The dotted curve represents a calculation that includes
only conventional nucleonic degrees of freedom.
The agreement at lower energies is reasonable and expected
since these energies probe the long-range N-N interaction
that is well understood.
As the energy is increased, the shorter-ranged physics
become progressively more important, and the conventional
model deviates from the available data.
Similarly, for a fixed energy,
the conventional model
shows greater descrepancy as the
scattering angle increases,
corresponding to larger momentum transfers that
again probe shorter-ranged physics.
The dashed curve depicts a calculation
that includes
$\Delta$ isobar degrees of freedom.
As expected $\Delta$ isobar effects
increase with increasing energy or momentum transfer.
The value of the $\Delta$(1232) magnetic moment
is taken to be 4.0 $\mu_N$, consistent with the experimentally
determined value~\cite{GREC}.
The strengths of the $\pi$ and $\rho$ exchange potentials
are determined by the meson masses
and the $\Delta N \pi$ and $\Delta N \rho$ coupling constants
of reference ~\cite{NIEP}.
The value of the
${f^2_{\Delta N \rho}}\over {4\pi}$
coupling constant is not at all well
known~\cite{JAIN}
and may be
considerably smaller than 9.13, the value used in this study.
The sensitivity to $\rho$ exchange is indicated by the dash-dot
curve, where the $\rho$ exchange potentional has been
reduced by 50 percent, and by the solid curve where it has
been set to zero.
Consistent with the above momentum-range arguments,
the longer-ranged $\pi$ exchange contribution is more
important at lower energies than $\rho$ mediated effects which
dominate at higher energies.
The calculation with the $\rho $ exchange potential
set to zero best represents the data
and is consistent with a smaller
$\Delta N \rho $ coupling constant~\cite{JAIN}.

There is a limited amount of
$^{16}O(\gamma ,n)^{15}O$
data available in this energy range.
Angular distributions for this data, as well as the available
$^{16}O(\gamma ,p)^{15}N$ data
at the corresponding energies, are
shown in fig. 2
along with our calculations.
As in fig. 1, the dotted line represents the
conventional calculation involving only nucleonic
degrees of freedom.
The solid line is the calculation including
$\Delta$ degrees of freedom,
using the coupling constants of reference~\cite{NIEP},
with the $\rho $ exchange potential taken to be zero
as in figure 1.
The regularization parameter,
$\Lambda$,
was taken to be 1.2 GeV, which
is within the usually quoted range~\cite{NIEP}.
In principle the regularization should be independent
of $\Lambda$; however, for the large momentum
transfers considered here, we have found appreciable
numerical sensitivity.
The dashed curve is
for $\Lambda $ = 3.0 GeV with only
$\pi$ exchange.
Our calculation is stationary for this value of
$\Lambda$, becoming insensitive to small
changes
in this parameter.
The difference between the solid and dashed curves then
roughly indicates the uncertainty in our
calculation due to the regularization prescription.
This difference not only indicates a clear need for a better
regularization procedure, but also, and more importantly,
an exciting opportunity to study
short-range physics in a
many-body environment by using an
improved, more realistic interaction.

In summary, a key new result of this study is
that rigorous
conventional calculations involving only
nucleonic degrees of freedom
describe the available data remarkably well.
They are insufficient, however, to provide a complete
quantitative description
of
$^{16}O(\gamma ,n)$ and $^{16}O(\gamma ,p)$
data at intermediate energies
with the descrepancy
increasing with larger momentum transfer.
The inclusion of the $\Delta$(1232) resonance,
with its intrinsic short-ranged nature,
somewhat improves the description of the
available data,
but,
with the exception of the high energy $\Delta$ resonance region,
the effects are small and complicated by the
uncertainties in the short-ranged character of the interaction
and our regularization procedure.
This is in contrast to the results of refs.~\cite{LOND,CHEU}, which both find
significant contributions to the
$^{16}O(\gamma ,p)$
reaction,
although ref.~\cite{LOND} finds a much larger effect then ref.~\cite{CHEU}.
However, their calculations,
which are both direct knockout, omit what we find are
very important non-local microscopic elements.
In particular,
without charge exchange it is definitely not possible
for our model to reproduce the correct magnitude of
the
$^{16}O(\gamma ,n)$ data.

Perhaps the most important element not included in
this analysis
is meson exchange currents.
As is appropriately noted ~\cite{LOND},
at higher energies the photon probes the two-body
component of the current, and
exchange current effects should be
increasingly more important
as demonstrated ~\cite{MCDE} previously.
Reference ~\cite{MCDE} calculates significant
relativistic and exchange current effects, and finds that
both effects are necessary to describe the
$^{16}O(\gamma ,p)$ data.
However that same model, even with exchange currents, cannot
account ~\cite{BEIS}
for the
magnitude of the
$^{16}O(\gamma ,n)$ cross section.
The work of ref.~\cite{GARI} also  implies that meson exchange
currents play a major role in determining the
$^{16}O(\gamma ,p)$ reaction cross section.
Similar to our finding, and perhaps surprising,
this work also concludes that there
is no significant $\Delta$ isobar contribution.
While we agree with the approach of ref. ~\cite{GARI} for low energies,
the extension  to intermediate energies is
inappropriate due to the limitations of Siegert's theorem in
determining the exchange currents. Further, our calculations
clearly
indicate that magnetic multipoles, which are ignored in their
calculation, are not negligible.

The diverse results in the above investigations clearly
indicate that conclusions characterizing exchange current effects
can be very model dependent.
We submit that these differing findings are due to the nature
and degree of approximations, and that an accurate assessment of
exchange current effects requires a consistent, many-body framework.

A second major result of our work
is the description of
$\bf {both}$
$^{16}O(\gamma ,n)$ and $^{16}O(\gamma ,p)$
using the same microscopic model.
It should be emphasized that our conventional model
parameters were determined at lower energies by
independently fitting the
$^{16}O(\gamma ,p)$ giant dipole resonance reaction,
and therefore,
with the exception of variations in the $\Lambda$
potential parameters to document regularization sensitivity,
all of our
$^{16}O(\gamma ,N)$ medium energy
calculations are
actually model predictions.
While our calculations document the importance
of the $\Delta$
resonance, it should be noted that there
is strong sensitivity to the regularization
parameter $\Lambda$, indicating
not only the need for an improved model interaction, but also,
and more importantly, the opportunity for studying
interesting short range physics including baryon repulsion.
Efforts are presently underway
to incorporate more sophisticated
short-ranged
$N-N$ and $N-\Delta$ interactions
and meson exchange currents which
we also expect to be important.

This work was supported by the U.S. Department
of Energy, grant DE-FG05-88ER40461,
and the North Carolina Supercomputing Center.

\vfill\eject


\vfill\eject

\begin{center}
\section*{Figure Captions}
\end{center}
\begin{description}
\item{Fig.~1}
Laboratory Energy Spectra for the reaction
$^{16}O(\gamma ,p)^{15}N$.
Theory: conventional
calculation with only nucleonic degrees of freedom (dotted);
calculation with full
contribution from $\Delta$ degrees of freedom(dashed);
calculation with $\rho $ exchange potential reduced by 50 (dot-dash)
and 100 percent(full).
Data: squares are from ref.~\cite{FIND}. Circles are from
ref.~\cite{LEIT}.
The $90^o$ and $135^o$ spectra are scaled by $10^{-1}$ and
$10^{-2}$ respectively.

\item{Fig.~2}
$^{16}O(\gamma ,n)^{15}O$
and
$^{16}O(\gamma ,p)^{15}N$
laboratory angular distributions for
$E_\gamma$=150,
200,
250 MeV.
Theory: Conventional calculation(dotted);
calculation with $\rho $ exchange potential
set to zero and $\Lambda$ = 1.2 GeV(full),
3.0 GeV (dashed).
Data: $(\gamma ,n)$ - ref.~\cite{BEIS};
$(\gamma ,p)$ - circles are from ref.~\cite{ADAM},
squares are from ref.~\cite{LEIT}.
The angular distributions for 150 and 200 MeV
are scaled by $10^4$ and $10^2$ respectively.

\end{description}

\end{document}